\documentstyle[12pt]{article}
\renewcommand
\baselinestretch{1.4}

\def\brr{\begin{array}}
\def\beq{\begin{equation}}
\def\ben{\begin{enumerate}}
\def\een{\end{enumerate}}
\def\err{\end{array}}
\def\eeq{\end{equation}}
\def\bea{\begin{eqnarray}}
\def\eea{\end{eqnarray}}
\def\bs{\bigskip}

\def\wt{\widetilde}

\def\nn{\nonumber}

\begin{document}

%\hfill UB-ECM-PF 93/

%\hfill HUPD-93-

\hfill March 1993

\vspace*{3mm}

\begin{center}

{\LARGE \bf
The renormalization structure and quantum equivalence of 2D dilaton
gravities}
\vspace{4mm}

\renewcommand
\baselinestretch{0.8}

{\sc E. Elizalde}\footnote{E-mail address: eli @ ebubecm1.bitnet}
\\
{\it Department E.C.M., Faculty of Physics, University of
Barcelona, \\
Diagonal 647, 08028 Barcelona, Spain} \\
{\sc S. Naftulin} \\
{\it Institute for Single Crystals, 60 Lenin Ave., 310141 Kharkov,
Ukraine} \\  and \\
{\sc S.D. Odintsov}\footnote{On  leave from
Tomsk Pedagogical Institute, 634041 Tomsk, Russia. E-mail address:
odintsov @ theo.phys.sci.hiroshima-u.ac.jp} \\ {\it Department of
Physics, Faculty of Science, Hiroshima University, \\
Higashi-Hiroshima 724, Japan}
\medskip

\renewcommand
\baselinestretch{1.4}

\vspace{5mm}

{\bf Abstract}
\end{center}

The one-loop effective action corresponding  the general model of
dilaton gravity given by the Lagrangian $L=-\sqrt{g} \left[
\frac{1}{2}Z(\Phi ) g^{\mu\nu} \partial_\mu \Phi \partial_\nu \Phi
+ C(\Phi ) R + V (\Phi )\right]$, where $Z(\Phi )$, $ C(\Phi )$ and
$V (\Phi )$ are arbitrary functions of the dilaton field, is found.
The question of the quantum equivalence of classically equivalent
dilaton gravities is studied. By specific calculation of explicit
examples it is shown that classically equivalent quantum gravities
are also perturbatively equivalent at the quantum level, but only
on-shell. The renormalization group equations for the generalized
effective couplings  $Z(\Phi )$, $ C(\Phi )$ and $V (\Phi )$ are
written. An analysis of the equations shows, in particular, that
the Callan-Giddings-Harvey-Strominger model is not a fixed point of
these equations.

\vspace{3mm}

%\noindent PACS: 04.50, 03.70, 11.17.

\newpage

\section{Introduction}

2D dilaton gravity constitutes a very nice example of a toy model
for 4D quantum gravity, a theory that has not been formulated yet
in a consistent way. The study of 2D dilaton gravity can throw new
light into some of the general properties of 4D gravity and, of
course, it may actually predict some new unexpected phenomena.

In particular, one of the topics which is currently under
discussion in 2D dilaton gravity (motivated by the recent
identification of black holes in string theories [1]) concerns the
{\it quantum} structure of 2D dilaton gravity and of 2D black holes
[1-9] (for a review and extended list of references see [6]). There
is the hope that the longstanding mystery concerning the Hawking
evaporation of black holes [11] will be solved in a theory of 2D
gravity with matter (see [2-4,6] for a discussion of recent
progress on this point).

Actually, two different models of 2D dilaton gravity have been
discussed in the literature. A very general model of such theory,
which is multiplicatively renormalizable (first ref. of [7],[8] and
[15]) is given by the following action
\begin{equation}
S=-\int d^{2}x \,\sqrt{g}\left[ {1\over 2} Z( \Phi ) g^{\mu
\nu}\nabla_\mu \Phi  \nabla_\nu \Phi + C(\Phi ) R +V(\Phi )\right],
\label{so}
\end{equation}
where  $Z(\Phi )$, $ C(\Phi )$ and $V (\Phi )$ are some functions
of the dilaton field $\Phi$. This theory can be considered as a
kind of $\sigma$-model or one-dimensional string (see, for example
[12]). A popular choice of these functions ---which corresponds to
the model of Callan, Giddings, Harvey and
Strominger (CGHS) [2] is the following
\[
Z(\Phi )= 8e^{-2\Phi}, \ \ \ \  C(\Phi )= e^{-2\Phi}, \ \ \ \ V
(\Phi )= 4 \lambda^2 e^{-2\Phi}.
\]
One can also consider other choices for the functions in (1).
Some of them lead to nonsingular theories [8].

Let us now discuss an important problem which appears in the
different  2D quantum gravities based on the action (1). The idea
is the following. One can start from some particular model of the
family (1), with specified functions   $Z(\Phi )=Z_1(\Phi )$, $
C(\Phi )= C_1(\Phi )$ and $V (\Phi )=V_1 (\Phi )$ (for instance, a
model motivated by string theory), or from a different one (let us
say coming from a $\sigma$ model as the one above), with
corresponding functions  $Z_2(\Phi )$, $ C_2(\Phi )$ and $V_2 (\Phi
)$. Before going on, we shall perform the transformation
\beq
 g_{\mu  \nu} \longrightarrow e^{-2\rho (\varphi)}  \wt{g}_{\mu
\nu}, \ \ \ \ \Phi \longrightarrow f(\varphi ),
\eeq
and work with the new variables, what renders some expressions more
simple. So the formal setting is that  we have two theories of the
class (1) characterized by two different sets of functions:  $\{
Z_1(\Phi ), C_1(\Phi ),V_1 (\Phi )\}$ and  $\{
Z_2(\Phi ), C_2(\Phi ),V_2 (\Phi )\}$. Of course, at the classical
level these two theories are equivalent and lead to the same
classical physics. Now, the natural question is if this
equivalence will be maintained at the quantum level. Are any two
classically equivalent theories of the family (1) also quantum
equivalent? If the answer is {\it no}, then the physics of these
two theories will be different. (This is certainly an important question
which also showed up in the early days of string theories and
concerned the classical, semiclassical and quantum equivalence of
the different string models.)\,  In fact there are some indications
[4] that classical equivalence does not carry over to the quantum
level.

The present work is devoted to the study of the effective action
and to the question of quantum equivalence in 2D gravity within the
covariant perturbative approach. We will calculate the one-loop
effective action and shall show that, in general, classically
equivalent 2D dilatonic gravities are not quantum equivalent
off-shell. The paper is organized as follows. In the next section
we discuss a popular model of dilatonic gravity. We calculate the
one-loop effective action in two minimal gauges. After that, the
one-loop effective action in a classically equivalent version of
the same dilaton gravity in one of the two gauges considered
(conveniently transformed) is obtained. Comparison of the results
will show, in fact, that these two models of dilaton gravity that
are clasically equivalent are also quantum equivalent on-shell. In
sect. 3 the one-loop effective action for the general model (1) of
dilaton gravity interacting with a Maxwell field is found. It is
shown that this effective action is given by a  total derivative
term on shell (finiteness property). Sect. 4 is devoted to the
renormalization group analysis of the generalized effective
couplings  $Z(\Phi )$, $ C(\Phi )$ and $V (\Phi )$. Some variants
of fixed points for renormalization group $\beta$-functions are
presented. Sect. 5 is devoted to conclusions. We also make there
some final remarks on our results.
In an appendix we show that the results of sect. 3 provide
also the one-loop counterterms corresponding to 2D $R^2$-gravity.
 \bs

\section{One-loop effective action and quantum equivalence of
dilatonic gravities: an example}

In this section we will study the popular version of 2D dilatonic
gravity which is given by the action
\begin{equation}
S_1=-\int d^{2}x \,\sqrt{g}\left[ {1\over 2} g^{\mu  \nu}\partial_\mu
\Phi  \partial_\nu \Phi + CR \Phi +V(\Phi )\right],
\label{so1}
\end{equation}
where $C$ is a positive constant and $V(\Phi )$ an arbitrary
function. The one-loop renormalization of the theory (3) has been
performed in refs. [7] in different covariant gauges.

Let us here briefly summarize the  results of refs. [7,14]
concerning the one-loop effective action. We use the background
field method
\begin{equation}
g_{\mu\nu} \longrightarrow \bar{g}_{\mu\nu} =g_{\mu\nu}
+h_{\mu\nu}, \ \ \ \ \
\Phi \longrightarrow \bar{\Phi} = \Phi +
\varphi,
\end{equation}
where $h_{\mu\nu}$ and $\varphi$ are the quantum fields. The
simplest minimal covariant gauge is given by
\begin{equation}
S_{GF}=-{1\over 2}\int d^2x\,c_{\mu\nu}\, \chi^\mu\chi^\nu,
%\label{}
\end{equation}
where
\beq
c_{\mu\nu} = - C \Phi \sqrt{g}\, g_{\mu\nu}, \ \ \ \ \chi^\mu = -
\nabla^\nu \bar{h}^\mu_{\ \nu} + \frac{1}{\Phi} \nabla^\mu \varphi,
\eeq
and  $\bar{h}_{\mu\nu}=h_{\mu\nu}-\frac{1}{2} g_{\mu\nu} h$. The
divergences of the one-loop effective action (including all surface
terms) have been calculated in ref. [14]
\beq
\Gamma_{div}=-{1\over 2\epsilon}\int d^2 x\,\sqrt g\,
\Biggl[4R+{2\over C\Phi}V+{2\over
C}V'+\biggl({1\over \Phi}-{1\over C} \biggr) (\Delta \Phi )- {3\over
\Phi^2}(\nabla^\lambda\Phi)
 (\nabla_\lambda\Phi)\Biggr].
%\label{}
\eeq
Notice that in the first and third references of [7] such
calculation has been done without taking into account the surface
terms and also, that the theory (3) is {\it not} invariant under
the change $\Phi \rightarrow  - \Phi$ (no $Z_2$ symmetry). Hence,
the classical restriction $\Phi \geq 0$ seems reasonable here (see
the paper by Hamada and Tsuchiya in [4]). This renders the
discussion of the quantum dynamics of black holes quite difficult
(usually, however, this restriction has been ignored).

Let us now make in the theory (3) the field transformation
\beq
\Psi^2=\frac{C}{\gamma} \Phi, \ \ \ \  g_{\mu  \nu} \longrightarrow
e^{-2\rho} \,  \wt{g}_{\mu \nu},
\eeq
where \[ \gamma >0, \ \ \ \rho = \frac{\gamma \Psi^2}{4C^2} -
\frac{1}{8\gamma} \ln \Psi. \] Then, action (3) becomes
\begin{equation}
S_2=-\int d^{2}x \,\sqrt{\wt{g}}\left[ {1\over 2} \wt{g}^{\mu
\nu}\partial_\mu
\Psi  \partial_\nu \Psi + \gamma \wt{R} \Psi^2 + U (\Psi)  \right],
\label{so2}
\end{equation}
where we have defined $ U (\Psi)\equiv e^{-2\rho} V(\Phi (\Psi) )$
and dropped off a total derivative term. The actions $S_1$ and $S_2$
(eqs. (3) and (9), respectively) belong to the same class (1) but
are parametrized through different triplets of functions $\{ Z, C,
V\}$. They are classically equivalent and lead to the same
classical physics.

We shall now investigate the one-loop effective action for the
theory (9). The calculation will be done in the same gauge (5)-(6),
that is also to be transformed in accordance with (8). So, the
natural prescription is
\ben
\item For the background fields we shall make the transformation
(8), where $\Psi$ and $\wt{g}_{\mu\nu}$ will be now the background
fields of the theory (9).
\item The quantum fields will be transformed according to the first
order Taylor expansion of eq. (8), that is
\beq
\varphi \longrightarrow \frac{2\gamma}{C}\, \Psi \, \eta, \ \ \
h_{\mu\nu}  \longrightarrow e^{-2\rho (\Psi )} \left[
\wt{h}_{\mu\nu}+ \left( \frac{1}{4\gamma\Psi}- \frac{\gamma}{C^2}
\Psi \right)\wt{g}_{\mu\nu} \eta \right].
\eeq
\een
We should recall now that in the background field method for
the theory (9),
\beq
\wt{g}_{\mu\nu} \longrightarrow \wt{g}_{\mu\nu} +\wt{h}_{\mu\nu},
\ \ \ \ \Psi  \longrightarrow \Psi + \eta,
\eeq
where $\wt{h}_{\mu\nu}$ and $\eta$ are the quantum fields. Notice
also that from the functional integral point of view the change of
variables (10) is local.

Taking into account all the remarks above, we get the following
covariant gauge for the theory (9):
\begin{equation}
S_{GF}=-{1\over 2}\int d^2x\,c_{\mu\nu}\, \chi^\mu\chi^\nu,
%\label{}
\end{equation}
where
\bea
c_{\mu\nu} &=& -\gamma \Psi^2 \sqrt{g}\, g_{\mu\nu}, \\
\chi^\mu &=& -\nabla^\nu \bar{h}_{\mu\nu} +  \left(
\frac{\gamma}{C^2} \Psi - \frac{1}{4\gamma\Psi} \right)
(\nabla^\nu \Psi) \bar{h}^\mu_{\ \nu} + \frac{2}{\Psi}
\nabla^\mu \eta + \frac{2}{\Psi^2}(\nabla^\mu \Psi) \eta. \nn
\eea
In order to simplify notation,  in what follows we shall suppress
tildas over $g_{\mu\nu}$ and $h_{\mu\nu}$. Since we are working
with theory (9) only, this should not lead to any confusion. On the
other hand, it is no surprise at all that the gauge (12) is minimal
again (namely a minimal gauge is mapped into another minimal
gauge).

For the calculation of the one-loop effective action we will use
the standard technique
\begin{eqnarray}
\Gamma _{div}&=& \frac{i}{2} \mbox{Tr}\,  \ln \left.
\hat{H}\right|_{div}
= \frac{i}{2} \mbox{Tr}\,  \ln \left.
 (\hat{1} \Delta+2 \hat{E}^{\lambda}
\nabla_{\lambda} +\hat{\Pi} )
\right|_{div} \nn\\
& =& {1\over 2\epsilon}\int\,d^2x\, \sqrt{g} \, \mbox{Tr}\, \left[
\hat{\Pi}+{R\over6}\hat1-
\hat{E}^\lambda\hat{E}_\lambda-\nabla_\lambda\hat{E}^\lambda
\right],
\label{div-part}
\end{eqnarray}
where $\epsilon = 2\pi (n-2)$ and dimensional regularization has
been used.

The total quadratic expansion of the action (9) with gauge-fixing
term (12) can be written as follows
\begin{equation}
-\frac{1}{2}\varphi^i\hat{H}\varphi^j\equiv
-\frac{1}{2}\varphi^i\left[
-\hat{K}_{ij}\Delta+\hat{L}_{\lambda,
ij}\nabla^\lambda+\hat{P}_{ij}\right] \varphi^j,
\end{equation}
where $\varphi^i \equiv \{ \eta, h, \bar{h}_{\mu\nu} \}$ and
\beq
\hat{K}_{ij} = \left( \brr{ccc} 1-4\gamma & \gamma \Psi & 0 \\
 \gamma \Psi & 0 & 0 \\
0 & 0 & - \frac{1}{2} \gamma \Psi^2 P^{\mu\nu, \alpha\beta} \err
\right), \eeq
where $ P^{\mu\nu, \alpha\beta} = \delta^{\mu\nu, \alpha\beta}-
\frac{1}{2} g^{\mu\nu}  g^{\alpha\beta}$ and
\begin{eqnarray}
&&\hat{L}^\lambda_{11}=-\frac{8\gamma} {\Psi}  (\nabla^\lambda
\Psi), \ \ \ \ \hat{L}^\lambda_{12}=-\hat{L}^\lambda_{21}=\gamma
(\nabla^\lambda \Psi), \nn \\
&&\hat{L}^\lambda_{13}=-\hat{L}^\lambda_{31}= \left( \frac{1}{2} +
\frac{2\gamma^2\Psi^2}{C^2} \right)  (\nabla_\omega
\Psi)P^{\alpha\beta, \lambda \omega}, \nn \\
&&\hat{L}^\lambda_{22}=-\gamma \Psi   (\nabla^\lambda \Psi), \ \ \
\ \hat{L}^\lambda_{23}=-\hat{L}^\lambda_{32}= \gamma \Psi
(\nabla_\omega \Psi)P^{\alpha\beta, \lambda \omega}, \nn \\
&&\hat{L}^\lambda_{33}= \left( \frac{\gamma^2\Psi^3}{C^2} +2\gamma
\Psi- \frac{\Psi}{4} \right)  (\nabla^\omega \Psi) (P^{\mu\nu}_{
\omega \kappa} P^{\alpha\beta, \lambda \kappa}-   P^{\alpha\beta}_{
\omega \kappa} P^{\mu\nu, \lambda \kappa} ) -3 \gamma \Psi
(\nabla^\lambda \Psi)P^{\mu\nu,\alpha\beta}, \nn \\
&&\hat{P}_{12}=\hat{P}_{21}=\frac{1}{2}U', \ \ \ \ \hat{P}_{22}=0,
\nn \\
&&\hat{P}_{33}=-{1\over 2}\left[\gamma \Psi^2R+ U+ {1\over2}
(\nabla_\lambda \Psi) (\nabla^\lambda \Psi) \right]
P^{\mu\nu,\alpha\beta} \nn \\
&& + \left[ {5\over 4} - 4\gamma - {1\over 16\gamma} -
\frac{3\gamma^2\Psi^2}{C^2}+ \frac{\gamma\Psi^2}{2C^2}-
\frac{\gamma^3\Psi^4}{C^4}\right](\nabla_\lambda \Psi)
(\nabla^\omega \Psi)P^{\mu\nu}_{
\omega \kappa} P^{\alpha\beta, \lambda \kappa}, \nn \\
&& + \left[ {\Psi\over 4} - 4\gamma\Psi -\frac{\gamma^2\Psi^3}{C^2}
\right](\nabla^\omega\nabla_\lambda \Psi)P^{\mu\nu}_{
\omega \kappa} P^{\alpha\beta, \lambda \kappa}.
\eea
It is easy to see that the operator $\hat{H}$ is not uniquely
defined, since arbitrary integrations by parts can be performed. In
order to eliminate this possibility and end up with a uniquely defined
hermitean operator, the doubling trick of 't Hooft and Veltman [13]
is very useful. Applying it amounts to doing the following
redefinitions in  $\hat{H}$ (15):
\begin{eqnarray}
&\hat{H}\to\hat{H}'=-\hat{K}\Delta+\hat{L}'_\lambda\nabla^\lambda
+\hat{P}',  &\cr
&\hat{L}'_{\lambda}={1\over2}\big(\hat{L}_\lambda-
\hat{L}^T_\lambda\big)
                    -\nabla ^\lambda\hat{K} ,
 &\cr
&\hat{P'}={1\over2}\big(\hat{P}+\hat{P}^T\big)-
{1\over2}\nabla^\lambda
          \hat{L}^T_\lambda-{1\over2}\Delta\hat{K},
    &
\label{corrected-matrices}
\end{eqnarray}

Introducing the notations $\hat{E}^\lambda=-(1/2)\hat{K}^{-
1}\hat{L}^{'\lambda}$ and  $\hat{\Pi}=-\hat{K}^{-1}\hat{P}'$, the
operator $\hat{H}'$ can be put in the form
\begin{equation}
\hat{H}'=-\hat{K} (\hat{1} \Delta+2 \hat{E}^{\lambda}
\nabla_{\lambda} +\hat{\Pi} ).
\end{equation}
where
\bea
&& (\hat{E}^\lambda)_1^1 =\frac{1}{\Psi} (\nabla^\lambda \Psi), \
\ \ (\hat{E}^\lambda)_2^1 =0, \ \ \ (\hat{E}^\lambda)_3^1
=-\frac{1}{2} (\nabla_\omega \Psi)  P^{\alpha\beta, \lambda \omega},
\nn\\ && (\hat{E}^\lambda)_1^2 =\frac{4\gamma -1}{\gamma\Psi^2}
(\nabla^\lambda \Psi), \ \ \
 (\hat{E}^\lambda)_2^2 =0, \ \ \  (\hat{E}^\lambda)_3^2
=\left(\frac{1}{4\gamma \Psi}- \frac{2}{\Psi}-
\frac{\gamma\Psi}{C^2} \right) (\nabla_\omega \Psi)
P^{\alpha\beta, \lambda \omega},  \nn \\ &&  (\hat{E}^\lambda)_1^3
=-\left(\frac{1}{2\gamma \Psi^2}+ \frac{2\gamma}{C^2} \right)
(\nabla_\omega \Psi)  P_{\rho\sigma}^{\lambda \omega}, \ \ \
(\hat{E}^\lambda)_2^3 =-\frac{1}{\Psi} (\nabla_\omega \Psi)
P_{\rho\sigma}^{\lambda \omega}, \nn \\
&&   (\hat{E}^\lambda)_3^3 =\left(\frac{\gamma\Psi}{C^2}+
\frac{2}{\Psi}- \frac{1}{4\gamma\Psi} \right) (\nabla^\omega \Psi)
( P_{\rho\sigma,\omega\kappa} P^{\alpha\beta, \lambda \kappa}-
P^{\lambda\kappa}_{\rho\sigma}
P^{\alpha\beta}_{\omega\kappa})+\frac{1}{\Psi} (\nabla^\lambda
\Psi)  P^{\alpha\beta}_{\rho \sigma}, \nn \\
&& (\hat{\Pi})_1^1 = \frac{1}{\Psi}\Delta \Psi
-\frac{1}{2\gamma\Psi} U', \ \ \  (\hat{\Pi})_2^2 =-
\frac{1}{2\gamma\Psi} U' +\frac{1-4\gamma}{2\gamma\Psi}\Delta \Psi
 +\frac{1-4\gamma}{2\gamma\Psi^2}  (\nabla^\lambda \Psi)
(\nabla_\lambda \Psi), \nn \\
&& (\hat{\Pi})_3^3 =\left[ \frac{8\gamma-1}{2\gamma\Psi^2}
(\nabla^\lambda \Psi)  (\nabla_\lambda \Psi) +\frac{4}{\Psi}\Delta
\Psi -R -\frac{1}{\gamma\Psi^2} U\right]  P^{\alpha\beta}_{\rho
\sigma}, \nn \\
&& +\left[ \frac{1}{2\gamma\Psi}-\frac{8}{\Psi}-
\frac{2\gamma\Psi}{C^2} \right] (\nabla^\omega\nabla_\lambda \Psi)
P_{\rho\sigma,\omega\kappa} P^{\alpha\beta, \lambda \kappa} \\
&& +\left[ \frac{20\gamma-64 \gamma^2-1}{8\gamma^2\Psi^2}+\frac{1-
6\gamma}{C^2}-\frac{2\gamma^2\Psi^2}{C^4} \right] (\nabla^\omega
\Psi)(\nabla_\lambda \Psi) P_{\rho\sigma,\omega\kappa}
P^{\alpha\beta, \lambda \kappa} \nn
\eea
Notice that we do not need the off-diagonal components of
$\hat{\Pi}$.

Evaluating the functional traces in (14), we get
\bea
\Gamma_{2-div}&=&-{1\over 2\epsilon}\int d^2 x\,\sqrt g\,
\left[\frac{4}{3}R+{2\over \gamma\Psi^2}U+{1\over
\gamma\Psi}U'+\left({4\gamma-1\over \Psi}+{2\gamma\Psi\over C^2}
\right)(\Delta \Psi) \right.\nn \\
&+& \left. \left( {2\gamma\over C^2}-  {2\over
\Psi^2}\right) (\nabla^\lambda\Psi)
 (\nabla_\lambda\Psi)\right].
%\label{}
\eea
The ghost operator turns out to be
\bea
\hat{\cal M}_\nu^\mu &=& g_\nu^\mu \Delta + \left( {1\over
4\gamma\Psi} -  {\gamma \Psi\over
C^2}\right) (\nabla^\sigma\Psi)
\nabla_\lambda \left( g_\nu^\mu g_\sigma^\lambda+ g_{\sigma\nu}
g^{\lambda\mu}- g_\sigma^\mu g_\nu^\lambda \right) \nn \\
&-&\frac{2}{\Psi} (\nabla_\nu \Psi) \nabla^\mu -\frac{2}{\Psi}
(\nabla^\mu \nabla_\nu \Psi)-\frac{2}{\Psi^2} (\nabla_\nu \Psi)(
\nabla^\mu \Psi) + R_\nu^\mu,
\eea
and the divergent ghost contribution can be written as
\bea
\Gamma_{ghost-div}&=&-{1\over 2\epsilon}\int d^2 x\,\sqrt g\,
\left[\frac{8}{3}R+\left({2\gamma\Psi\over C^2}-{1+4\gamma\over
2\gamma\Psi}\right) (\Delta \Psi)\right.\nn \\ &+& \left.
\left({2\gamma\over C^2}+{1-16\gamma\over 2\gamma\Psi^2} \right)
(\nabla^\lambda\Psi)
 (\nabla_\lambda\Psi)\right].
%\label{}
\eea
The total one-loop divergence, $\Gamma_{div} =\Gamma_{2-
div}+\Gamma_{ghost-div}$, is given by
\bea
\Gamma_{div}&=&-{1\over 2\epsilon}\int d^2 x\,\sqrt g\,
\left[4R+{2\over \gamma\Psi^2}U+{1\over
\gamma\Psi}U'+\biggl({4\gamma-3\over 2\gamma\Psi}+{4\gamma\Psi\over
C^2} \biggr)(\Delta \Psi) \right.\nn \\
&+& \left. \left( {4\gamma\over C^2}+  {1-20\gamma\over
2\gamma\Psi^2}\right) (\nabla^\lambda\Psi)
 (\nabla_\lambda\Psi)\right].
%\label{}
\eea

We must observe that in this final formula all surface terms have been
kept. A few remarks are in order.
First of all, let us do an integration by parts in (24) and drop
the surface terms. We get
\beq
\Gamma_{div}=-{1\over 2\epsilon}\int d^2 x\,\sqrt g\,
\left[{2\over \gamma\Psi^2}U+{1\over
\gamma\Psi}U'-{1+8\gamma\over \gamma\Psi^2}g^{\mu\nu}
(\nabla_\mu\Psi)
 (\nabla_\nu\Psi)\right].
%\label{}
\eeq

The one-loop renormalized action is given by
\beq
S_R = S-\Gamma_{div}.
\eeq
Making use of the renormalization transformation
\beq
g_{\mu\nu} = \exp \left({1+8\gamma\over
8\epsilon \gamma^2\Psi^2} \right) g_{\mu\nu}^R,
\eeq
where $g_{\mu\nu}^R$ is the renormalized metric, we get the
one-loop renormalized action
\beq
S_R=-\int d^2 x\,\sqrt{g_R} \,
\Biggl[ \frac{1}{2} g^{\mu\nu}_R \partial_\mu \Psi \partial_\nu
\Psi +\gamma \Psi^2R_R +U +{U\over
8\gamma^2 \epsilon \Psi^2}-{U'\over 2\epsilon\gamma\Psi}\Biggr].
%\label{}
\eeq

It follows from (28) that the theory under discussion is one-loop
off-shell renormalizable in the usual sense for the following
choice of $U$:
\beq
U= \exp \left( \frac{1}{4\gamma} \ln \Psi - a_1\gamma \Psi^2 +a_2
\right),
\eeq
where $a_1$ and $a_2$ are arbitrary constants.

Let us now discuss the on-shell limit of $\Gamma_{div}$, eq. (24).
Keeping all the surface terms and using the classical field
equations resulting from the action (9), namely
\beq
-\Delta \Psi +2\gamma R \Psi +U' =0, \ \ \ \ \ -\gamma \Delta
\Psi^2+U=0,
\eeq
we obtain
\beq
\Gamma_{div}^{on-shell}=-{1\over 2\epsilon}\int d^2 x\,\sqrt g\,
\Biggl[ 2R+ \Delta \left(\frac{12\gamma -1}{2\gamma} \ln \Psi
+{2\gamma\over C^2} \Psi^2 \right) \Biggr].
%\label{}
\eeq
Notice that this is a total derivative.

As a second example of the theory (9) we will consider the gauge
fixing action of the following form
\beq
S_{GF}= -{1\over 2}\int d^2 x\, c_{\mu\nu} \chi^\mu \chi^\nu,
\eeq
where $c_{\mu\nu} =-\gamma \Psi^2\sqrt{ g} \, g_{\mu\nu}$ and $ \chi^\mu
=-\nabla^\nu \bar{h}^\mu_\nu +\frac{2}{\Psi} \nabla^\mu \eta$.
Repeating the calculation above with this new gauge choice, one finds
the following final result for the one-loop effective action:
\beq
\Gamma_{div}=-{1\over 2\epsilon}\int d^2 x\,\sqrt g\,
\Biggl[ 4R+ {2\over \gamma\Psi^2}U+{1\over
\gamma\Psi}U'+{4\gamma-1 \over 2\gamma \Psi} \Delta \Psi-
{1+20\gamma\over 2\gamma\Psi^2} (\nabla^\lambda\Psi)
 (\nabla_\lambda\Psi)\Biggr].
%\label{}
\eeq
As we can see, $\Gamma_{div}$, eq. (33), differs from
$\Gamma_{div}$, eq. (24), in surface terms (a total derivative).
After integration by parts, eq. (33) becomes eq. (25). Hence, the
off-shell one-loop renormalization is the same in both gauges (32)
and (5). However, on shell  $\Gamma_{div}$ in eq. (33), {\it
differs} from eq. (31) in some total derivative terms. Summing up,
we see that the on-shell effective action in both gauges is given
by surface divergences only (finiteness of the $S$ matrix), but
these terms depend yet on the choice of gauge condition.

We are now going to investigate the theory (3) in the variables
(8). Transforming   $\Gamma_{div}$, eq. (7), to the new variables
(8), we get
\bea
\Gamma_{div}&=&-{1\over 2\epsilon}\int d^2 x\,\sqrt g\,
\left[ 4R+ {2\over \gamma\Psi^2}U+e^{-2\rho (\Psi)}{2\over
\gamma} \, \frac{\partial V(\Psi)}{\partial \Psi}\right.\nn \\
&+&\left. \left( {4\gamma-2 \over 2\gamma \Psi}+ {2\gamma\Psi \over
C^2}\right) \Delta \Psi+ \left( {2\gamma \over C^2}+{2-
20\gamma\over 2\gamma\Psi^2}\right) (\nabla^\lambda\Psi)
(\nabla_\lambda\Psi)\right],
%\label{}
\eea
where all surface terms have been kept.
As we see, there is no perturbative quantum equivalence between the
two classically equivalent dilaton gravities (3) and (9). This
result is a clear confirmation of the preliminary conclusions  in
ref. [4]. In fact, the one-loop effective action (34) which comes
from action (3) in the gauge (5) does {\it not} coincide with
$\Gamma_{div}$, eq. (24), which is obtained when starting from the
classically equivalent theory (9) in the gauge (12) ---that can be
made correspond with the gauge (5). There are, however, some
similarities between the two results.

In particular, the theory (3) in the gauge (5) is one-loop
off-shell multiplicatively renormalizable
 in the usual sense for the
following potential [7]:
\beq
V (\Phi) = e^{\alpha \Phi} + \Lambda,
\eeq
where $\alpha$ and $\Lambda$ are arbitrary constants. Making use of
the transformation (8) we obtain
\beq
U (\Psi) = e^{-2\rho(\Psi )}V (\Psi) = \exp \left( \frac{1}{4\gamma}
\ln \Psi-\frac{\gamma \Psi^2}{4C^2} \right) \left( e^{\ln \Lambda}
+ e^{\alpha\gamma \Psi^2/C} \right).
\eeq
This function $U$ belongs to the same class as the $U$ of eq. (29).
We thus see from our example that two classically equivalent
dilaton gravities  lead to the same class of multiplicatively
renormalizable potentials (which are of Liouville form).

Let us now consider the on-shell effective action of the theory
(3). Using classical field equations and keeping all the total
derivative terms in (7), we get
\beq
\Gamma_{div}^{on-shell}=-{1\over 2\epsilon}\int d^2 x\,\sqrt g\,
\Biggl[ 2R+ \Delta \left(\frac{1}{C}  \Phi +3\ln \Phi
\right)\Biggr].
%\label{}
\eeq
Transforming the variables in (37) according to the change (8), se
obtain
\beq
\Gamma_{div}^{on-shell}=-{1\over 2\epsilon}\int d^2 x\,\sqrt g\,
\Biggl[ 2R+ \Delta \left(\frac{2\gamma\Psi^2}{C^2} +\frac{12\gamma-
1}{2\gamma} \ln \Psi \right)\Biggr].
%\label{}
\eeq
This expression coincides completely with eq. (31).

{}From the discussion above, we conclude that there is on-shell
perturbative quantum equivalence for the classically equivalent
dilaton gravities (3) and (9). This statement is {\it not} true
off-shell, generally speaking, as follows from the preceding
analysis.
\bs

\section{The one-loop effective action in a general model of 2D
dilaton-Maxwell gravity}

In this section we will present an analysis of the one-loop
effective action in  a general model of 2D dilaton-Maxwell gravity.
It is defined by the following action
\begin{equation}
S=-\int d^{2}x \,\sqrt{g}\left[ {1\over 2} Z(\Phi ) g^{\mu
\nu}\partial_\mu
\Phi  \partial_\nu \Phi + C(\Phi )R +V(\Phi )+ \frac{1}{4} f(\Phi )
F_{\mu\nu}^2 \right].
\label{som}
\end{equation}
We shall use again the background field method. Fields will be
split according to (4) with the additional expression $A_\mu
\rightarrow A_\mu +Q_\mu$. In this section we will continue
denoting $C\equiv C (\Phi )$ (now $C$ is a function, not a constant
as in the previous section). The model with the action (39) is
connected (via some compactification) with the four-dimensional
Einstein-Maxwell theory, which admits charged black hole solutions
[10]. Particular cases of (39) describe the bosonic string and the
heterotic string effective actions, respectively.

The simplest minimal gauge fixing is
\beq
S_{GF}= -{1\over 2}\int d^2 x\,  \chi^A c_{AB} \chi^B,
\eeq
where $A \equiv \{ \mu, * \}$, $c_{\mu\nu} =-C\sqrt{ g} \, g_{\mu\nu}$,
 $ \chi^\mu =-\nabla_\nu \bar{h}^{\mu\nu} +\frac{C'}{C} \nabla^\mu
\varphi$, $C_*=\sqrt g f$, $\chi_* =-\nabla^\nu Q_\nu$. Let us
introduce $ \Phi^i = \{ Q_\mu, \varphi, h,  \bar{h}_{\mu\nu} \}$.

Now the quadratic contribution takes the usual form
\begin{equation}
S^{(2)}_{tot}=\int d^2x\, \sqrt{g} \,
\Phi^i\left(-\hat{K}\Delta+\hat{L}^\lambda
              \nabla_\lambda+\hat{P}\right)_{ij}\Phi^j,
\end{equation}
where
\begin{equation}
\hat{K}_{ij}=\pmatrix{ fg^{\mu\alpha} & 0 & 0 & 0 \cr
                              0 & Z-{{C'}^2\over C} & C'/2 & 0 \cr
                              0 & C'/2 & 0 & 0 \cr
                              0 & 0 & 0 &
-{C\over2}P^{\mu\nu,\alpha\beta} \cr}
\end{equation}
so that
\begin{equation}
\hat{K}^{-1}_{ij}=\pmatrix{ {1\over f}g_{\rho\mu} & 0 & 0 & 0 \cr
0 & 0 & 2/C' & 0
\cr
           0 & 2/C' & \left({4\over C}-{4Z\over{C'}^2}\right) & 0
\cr
           0 & 0 & 0 & -{2\over C}P_{\rho\sigma,\mu\nu} \cr} \ ,
\end{equation}
and the other essential matrix elements are:
\begin{eqnarray}
&&\hat{L}^\lambda_{11}=f'(\nabla^\alpha\Phi)g^{\mu\lambda}
-f'(\nabla^\mu\Phi)g^{\alpha\lambda}
                              -f'(\nabla^\lambda\Phi)g^{\mu\alpha}
\ ,  \cr\cr
&&\hat{L}^\lambda_{12}=-\hat{L}^\lambda_{21}
                                               =f'F^{\mu\lambda} \
,    \cr\cr
&&\hat{L}^\lambda_{13}=-\hat{L}^\lambda_{31}
                                     =-{1\over2}fF^{\mu\lambda} \
,   \cr\cr
&&\hat{L}^\lambda_{14}=-\hat{L}^\lambda_{41}
=fF^\lambda_{\phantom{\lambda}\omega}
P^{\alpha\beta,\mu\omega}-fF^\mu_{\phantom{\mu}\omega}P^{\alpha
\beta, \lambda\omega} \ , \cr\cr
&&\hat{L}^\lambda_{22}=\left(2{C'C''\over
C}-{{C'}^3\over C^2}+  Z'\right)(\nabla^\lambda\Phi)
\ ,  \cr\cr
&&\hat{L}^\lambda_{23}=-\hat{L}^\lambda_{32}
={1\over2}C''(\nabla^\lambda\Phi)\ ,\cr\cr
&&\hat{L}^\lambda_{24}=-\hat{L}^\lambda_{42}=
(Z-C'')(\nabla_\omega\Phi)P^{\alpha\beta,
                                      \lambda\omega } \ ,
\cr\cr
&&\hat{L}^\lambda_{33}=-{1\over2}C'(\nabla^\lambda\Phi)\ ,\cr\cr
&&\hat{L}^\lambda_{34}=-\hat{L}^\lambda_{43}=
{1\over2}C'(\nabla_\omega\Phi)P^{\alpha
                                      \beta,\lambda\omega} \ ,
\cr\cr
&&\hat{L}^\lambda_{44}=C'(\nabla^\omega\Phi)\left[P^{\mu
\nu}_{\omega\kappa}P^{\alpha\beta,\lambda
\kappa}-P^{\mu\nu,\lambda\kappa}P^{\alpha
\beta}_{\omega\kappa}\right]-
         {3\over2}C'(\nabla^\lambda\Phi)P^{\mu\nu,\alpha\beta} \ ,
    \cr\cr
&&\hat{P}_{11}=fR^{\mu\alpha} \ ,         \cr\cr
&&\hat{P}_{23}=\hat{P}_{32}={1\over2}V'-{1\over8} f'F^2 \
,   \cr\cr
&&\hat{P}_{33}={1\over8}fF^2 \ ,     \cr\cr
&&\hat{P}_{44}=\Bigl[\bigl(Z-2C''\bigr)(\nabla_\omega\Phi)
(\nabla^\lambda\Phi)-2C'(\nabla_\omega\nabla^\lambda\Phi)
+fF_{\omega\rho}F^{\lambda\rho}
\Bigr]P^{\mu\nu,\omega
         \kappa}P^{\alpha\beta}_{\lambda\kappa}    \cr\cr
&&\phantom{\hat{P}_{44}=}
-{1\over2}\left[CR+V+{1\over4}fF^2
+{1\over2}Z(\nabla^\lambda\Phi)
         (\nabla_\lambda\Phi)\right]P^{\mu\nu,\alpha\beta}   \cr\cr
&&\phantom{\hat{P}_{44}=}
+{1\over2}fF^{\omega\kappa}F^{\lambda
\rho}P^{\mu\nu}_{\omega\lambda}P^{\alpha\beta}_{\kappa\rho} \ .
\end{eqnarray}

The divergent part may be expressed in terms of the matrices
$\hat{E}^\lambda$ and $\hat{\Pi}$, in accordance with (19)
\begin{eqnarray}
&&\big(\hat{E}^\lambda\big)^{1}_{1}={f'\over2f}
\left[(\nabla^\lambda\Phi)
g^\alpha_\rho-(\nabla^\alpha\Phi)g^\lambda_
         \rho+(\nabla_\rho\Phi)g^{\alpha\lambda} \right] \ ,
\cr\cr
&&\big(\hat{E}^\lambda\big)^{1}_{2}=-{f'\over2f}F_\rho^{\phantom{
\rho}\lambda}
\ ,    \cr\cr
&&\big(\hat{E}^\lambda\big)^{1}_{3}={1\over4}F_\rho^{\phantom{\rho}
\lambda}
\ ,    \cr\cr
&&\big(\hat{E}^\lambda\big)^{1}_{4}={1\over2}F_{\rho\omega}
P^{\alpha\beta,
\lambda\omega}-{1\over2}F^{\lambda\omega}
                                    P^{\alpha\beta}_{\rho\omega} \
,    \cr\cr
&&\big(\hat{E}^\lambda\big)^{2}_{1}=-{f\over2C'}F^{\alpha\lambda}
\ ,   \cr\cr
&&\big(\hat{E}^\lambda\big)^{2}_{2}={C''\over
C'}(\nabla^\lambda\Phi)\ , \cr\cr
&&\big(\hat{E}^\lambda\big)^{2}_{3}=0 \ ,     \cr\cr
&&\big(\hat{E}^\lambda\big)^{2}_{4}=-{1\over
2}(\nabla_\omega\Phi)P^{\alpha
                                    \beta,\lambda\omega} \ ,
\cr\cr
&&\big(\hat{E}^\lambda\big)^{3}_{1}=\left({fZ\over{C'}^2}-{f\over
C}+{f'\over
                                    C'}\right)F^{\alpha\lambda} \
,     \cr\cr
&&\big(\hat{E}^\lambda\big)^{3}_{2}=\left({Z'\over
C'}-2{C''Z\over{C'}^2}+{{C'}^2\over
C^2}\right)(\nabla^\lambda\Phi)
                                                                \
,     \cr\cr
&&\big(\hat{E}^\lambda\big)^{3}_{3}=0 \ ,     \cr\cr
&&\big(\hat{E}^\lambda\big)^{3}_{4}=\left({C''\over C'}-{C'\over
C}\right)
(\nabla_\omega\Phi)P^{\alpha\beta,\lambda
                                    \omega} \ ,     \cr\cr
&&\big(\hat{E}^\lambda\big)^{4}_{1}={f\over
C}F^\alpha_{\phantom{\alpha}\omega}
P^{\lambda\omega}_{\rho\sigma}-{f\over C}
F^\lambda_{\phantom{\lambda}\omega}P^{\alpha\omega}_{\rho\sigma}
                                                            \ ,
  \cr\cr
&&\big(\hat{E}^\lambda\big)^{4}_{2}=\left({C''\over C}-{Z\over
C}\right)
(\nabla_\omega\Phi)P^{\lambda\omega}_{\rho
                                    \sigma} \ ,   \cr\cr
&&\big(\hat{E}^\lambda\big)^{4}_{3}=-{C'\over2C}
(\nabla_\omega\Phi)P^{\lambda \omega}_{\rho\sigma} \ ,
\cr\cr
&&\big(\hat{E}^\lambda\big)^{4}_{4}={C'\over
C}(\nabla^\omega\Phi)\Bigl[P_{\rho
\sigma,\omega\kappa}P^{\alpha\beta,\lambda
\kappa}-P^{\alpha\beta}_{\omega\kappa}P^{\lambda\kappa}_{\rho
\sigma}
\Bigr]+{C'\over2C}(\nabla^\lambda\Phi)P^{\alpha\beta}_{\rho\sigma}
                                                       \ ,
 \cr\cr
&&\hat{\Pi}^{1}_{1}=-R_\rho^\alpha \ ,   \cr\cr
&&\hat{\Pi}^{2}_{2}={C'''\over
C'}(\nabla^\lambda\Phi)(\nabla_\lambda\Phi)+
                    {C''\over C'}(\Delta\Phi)-{1\over
C'}V'+{f'\over 4C'}F^2
                                                            \ ,
\cr\cr
&&\hat{\Pi}^{3}_{3}=\left({C''Z\over{C'}^2}-{C''\over
C}\right)(\nabla^\lambda
                    \Phi)(\nabla_\lambda\Phi)+\left({Z\over
C'}-{C'\over C}
         \right)(\Delta\Phi)   \cr\cr
&&\phantom{\hat{\Pi}^{3}_{3}=} -{1\over
C'}V'+\left({fZ\over2{C'}^2}-{f\over2C}
                               +{f'\over4C'}\right)F^2 \ ,
\cr\cr
&&\hat{\Pi}^{4}_{4}=\biggl[\left(2{Z\over C}-4{C''\over
C}\right)(\nabla_\omega
                    \Phi)(\nabla^\lambda\Phi)-4{C'\over
C}(\nabla_\omega\nabla^\lambda\Phi)+{2f\over
C}F_{\omega\nu}F^{\lambda\nu}\biggr]P^{\omega
         \kappa}_{\rho\sigma}P^{\alpha\beta}_{\lambda\kappa}
\cr\cr
&&\phantom{\hat{\Pi}^{4}_{4}=} +\biggl[\left({2C''\over C}-{Z\over
2C}\right)
(\nabla^\lambda\Phi)(\nabla_\lambda\Phi)+2{C'
         \over C}(\Delta\Phi)-R-{1\over
C}V-{f\over4C}F^2\biggr]P^{\alpha\beta}_{\rho\sigma}      \cr\cr
&&\phantom{\hat{\Pi}^{4}_{4}=} +{f\over
C}F^{\omega\kappa}F^{\lambda\nu}P_{\rho
\sigma,\omega\lambda}P^{\alpha\beta}_{\kappa\nu} \ .
\end{eqnarray}
To obtain the divergent part, $\Gamma_{2-div}$, we have to evaluate
the functional traces of the matrices above as in eq. (14). Thus,
we get,
\begin{eqnarray}
\Gamma_{2-div}=-{1\over 2 \epsilon}\int d^2x\,\sqrt{g}\,\Biggl\{
                   2R+{2\over C}V+{2\over
C'}V'+\left({f'\over2C'}-{f\over2C}
                   \right)F^2     \qquad\cr\cr
+\left({f'\over f}+{2C'\over C}-{Z\over C'}\right)(\Delta\Phi)
\quad\cr\cr
+\left({f''\over f}-{{f'^2}\over
f^2}-{3{C'}^2\over2C^2}-{C''Z\over{C'}^2}
         \right)(\nabla^\lambda\Phi )(\nabla_\lambda\Phi)\Biggr\}
\ .
\end{eqnarray}

What is left to do is to calculate the divergent structure of the
both ghost operators: that corresponding to diffeomorphisms,
\begin{equation}
\widehat{\cal M}^\mu_\nu=g^\mu_\nu\Delta-{C'\over
C}(\nabla_\nu\Phi)\nabla ^\mu
                         -{C'\over
C}(\nabla^\mu\nabla_\nu\Phi)+R^\mu_\nu \ ,
\end{equation}
and the one corresponding to the Maxwell gauge transformations,
\begin{equation}
\widehat{\cal M}=\Delta \ .
\end{equation}
Hence,
\begin{equation}
\Gamma_{gh-div}=-{1\over 2\epsilon}\int d^2x\,\sqrt{g}\,\Biggl\{
                3R-{C'\over C}(\Delta\Phi)+\left({C''\over
C}-{3{C'}^2\over
2C^2}\right)(\nabla^\lambda\Phi)(\nabla_\lambda\Phi)\Biggr\}\ .
\end{equation}

The total divergent contribution is
\begin{eqnarray}
\Gamma_{div}=-{1\over 2\epsilon}\int
d^2x\,\sqrt{g}\,\Biggl\{5R+{2\over C}V
             +{2\over
C'}V'+\left({f'\over2C'}-{f\over2C}\right)F^2\qquad\cr\cr
+\left({f'\over f}+{C'\over C}-{Z\over C'}\right)(\Delta\Phi)
\quad\cr\cr
+\left({f''\over f}-{{f'^2}\over f^2}+{C''\over C}-3{{C'}^2\over
C^2}-
                 {C''Z\over{C'}^2}\right)(\nabla^\lambda\Phi
)(\nabla_\lambda
                 \Phi) \Biggr\} \ .
\end{eqnarray}
\medskip

Notice that from eq. (50) one can also get the one-loop effective
action for pure dilaton gravity with the action (1). The result is
\begin{eqnarray}
\Gamma_{div}=-{1\over 2\epsilon}\int
d^2x\,\sqrt{g}\,\Biggl\{4R+{2\over C}V
             +{2\over
C'}V'+\left({C'\over C}-{Z\over C'}\right) \Delta \Phi\qquad\cr\cr
+\left({C''\over C}-3{C'^2\over C^2}-{C''Z\over
C'^2}\right)(\nabla^\lambda\Phi
)(\nabla_\lambda\Phi) \Biggr\}.
\end{eqnarray}
In the expressions (50) and (51) for the effective action all
surface divergent terms have been kept.

Let us now discuss the on-shell limit of the effective action (50).
The classical field equations that we need are
\bea
\frac{\delta S}{\delta \Phi} &=& - \nabla_\nu
(Zg^{\mu\nu}\partial_\mu \Phi ) + \frac{1}{2} Z'(\nabla^\mu\Phi
)(\nabla_\mu\Phi) +C' R + V' + \frac{1}{4} f' F^2 =0, \nn \\
g^{\mu\nu}\frac{\delta S}{\delta g^{\mu\nu}} &=& -\Delta C +V -
\frac{1}{4} f F^2 =0.
\eea
Substituting eqs. (52) into the effective action (50) and keeping
all the surface counterterms, one obtains
\beq
\Gamma_{div}^{on-shell}=-{1\over 2\epsilon}\int d^2 x\,\sqrt g\,
\Biggl\{ 3R+ \Delta \left[ \ln (fC^3) \right] + \nabla^\lambda
\left[\frac{Z}{C'} \nabla_\lambda (\Phi ) \right]\Biggr\}.
%\label{}
\eeq
The theory with the action (39) is finite on-shell as well as the
dilaton gravities discussed in sect. 2. Hence, we can propose the
very plausible conjecture that dilaton gravities of the family (1),
if they are classically equivalent, are also quantum equivalent on
shell.
\bs

\section{Renormalization and renormalization group equations}

In this section we will study the renormalization structure and
renormalization group equations for dilaton gravity with the action
(1). One can discuss the renormalization of the metric as in sect.
2 and find the restrictions imposed by off-shell multiplicative
renormalizability  (in the usual sense) on the form of the
functions  $Z(\Phi )$, $ C(\Phi )$ and $V (\Phi )$. Instead of
doing this, we will not renormalize the fields but rather will
consider the functions $Z$, $C$ and $V$ as generalized effective
couplings (generalized renormalizability). Then the generalized
$\beta$-functions will be found and generalized renormalization
group equations will be generated.

Let us start from the theory defined by action (1). (For
simplicity, we shall not discuss the case of dilaton-Maxwell
gravity.) \, The general structure of renormalization for general
couplings is given by
\beq
Z_0 = (\mu^2)^{\epsilon'} \left[ Z+ \sum_{k=1}^\infty \frac{a_{kZ}
(Z,C,V)}{{\epsilon'}^k} \right],
\eeq
where $\epsilon'=n-2$, and similar expressions for $C$ and $V$. As
it follows from one-loop renormalization, eq. (51),
\beq
a_{1Z}=-\frac{Z'}{C'}+\frac{2{C'}^2}{C^2}+\frac{2C''Z}{{C'}^2}, \ \ \ \
\ \ a_{1C}=0, \ \ \ \ \ \  a_{1V}=-\frac{V}{C}-\frac{V'}{C'}.
\eeq
Now, the generalized one-loop $\beta$-functions are given by the
standard relations:
\beq
\beta_T = -a_{1T} + Z \frac{\delta a_{1T}}{\delta Z} + C
\frac{\delta a_{1T}}{\delta C} + V \frac{\delta a_{1T}}{\delta V},
\eeq
where $T\equiv \{ Z,C,V \}$. Using (55) and (56) we obtain
\bea
\beta_C &=& 0, \nn \\
\beta_V &=& \frac{V}{C}+\frac{V'}{C'}- \frac{V C''}{{C'}^2}-
\frac{CV''}{{C'}^2}+2\frac{CV'C''}{{C'}^3} \\
\beta_Z &=& \frac{Z'}{C'}+\frac{2{C'}^2}{C^2} -
\frac{4C''}{C}-\frac{ZC''}{{C'}^2}+3 \frac{CZ''}{{C'}^2}-
2\frac{CZ'C''}{{C'}^3}.  \nn
\eea
Of course, in the case that the theory includes a Maxwell sector,
similar $\beta$-functions for $f$ can be easily obtained too.

The renormalization group equations have the following form:
\beq
\frac{\partial C}{\partial t} = \beta_C, \ \ \ \ \ \ \ \frac{\partial
V}{\partial t} = \beta_V, \ \ \ \ \ \ \  \frac{\partial Z}{\partial t} =
\beta_Z,
\eeq
with $t=\ln \mu^2$. The system of equations (58) is very difficult
to solve. It depends not only on the scaling parameter $t$, as in
usual field theory, but also on some unknown functions of the field
variables. Moreover, nobody has any idea about the proper boundary
conditions (initial data) that the partial differential equation
system (58) should satisfy.

Notwithstanding that, we can get some useful information yet from
the renormalization group equations (58). In particular, we can look for
fixed points of this system (what does not at all involve the
knowledge of initial data). The equations they must satisfy are
\beq
\beta_C=0, \ \ \ \ \ \ \ \beta_V=0, \ \ \ \ \ \ \ \beta_Z=0.
\eeq
The system of differential equations (59) is still very
complicated. Nevertheless, some basic, particular solutions of the
same can be discovered. For example, motivated by the CGHS action
[2], we can look for fixed points of the following type:
\beq
C=e^{a_1\Phi}, \ \ \ \ V=e^{a_2\Phi}, \ \ \ \ Z=e^{a_3\Phi},
\eeq
where $a_1$, $a_2$ and $a_3$ are some constants. Substituting (60)
into eq. (59) we obtain the following solutions
\bea
&& a_1=0,  \ \ \ a_2=1, \ \ \ a_3 =- \frac{1}{3}; \nn \\
&& a_1=\frac{2}{3}, \ \ \ a_2=2, \ \ \ a_3 =\frac{1}{6} \left( 1\pm
\sqrt{19/3} \right); \ \ \ \mbox{etc.}
\eea
In the same way, different particular cases of fixed points can be
considered. For instance, for
\beq
C= \Phi^{\alpha_1}, \ \ \ \ V= \Phi^{\alpha_2}, \ \ \ \ Z=
\Phi^{\alpha_3},
\eeq
we get
\bea
&& \alpha_2 (\alpha_2 -4\alpha_1 +1)=0, \nn \\
&& 3\alpha_3^2-\alpha_3\alpha_1-\alpha_1^2-\alpha_3+\alpha_1=0.
\eea
Particular solutions are
\bea
&&\alpha_1=2, \ \ \ \alpha_2=0,  \ \ \ \alpha_3 =\frac{1}{2} \left(
3\pm \sqrt{33} \right); \nn \\
&&\alpha_1=2, \ \ \ \alpha_2=7,  \ \ \ \alpha_3 =\frac{1}{2} \left(
3\pm \sqrt{33} \right); \ \ \ \mbox{etc.}
\eea

Thus, we have shown that (at least in principle) one can find
ultraviolet stable fixed points for the generalized couplings
$Z(\Phi )$, $ C(\Phi )$ and $V (\Phi )$.
\bs

\section{Conclusions}

In summary, we have investigated in this paper
the one-loop renormalization structure of the general model of  2D
dilaton gravity (1).
The divergences of the one-loop effective action have been found.

The calculation of the one-loop effective action for two different
but classically equivalent dilaton gravities, eqs. (3) and (9),
respectively, in the same gauge (5) and (12), has shown that these
theories are quantum equivalent on shell. The one-loop on-shell
effective action is just given by surface terms, i.e., it is
finite. Since the on-shell effective action for the general model
of dilaton gravity (namely, including a Maxwell field) has the same
property, we have been led to conjecture that {\it all} classically
equivalent dilaton gravities are in fact quantum equivalent
on-shell.

Generalized renormalizability of the general model of dilaton
gravity has been discussed, and the corresponding generalized
$\beta$ functions have been found. The analysis of the
renormalization group equations yields some set of generalized
couplings $\{ Z,C,V \}$ which are fixed points of such equations.
It is interesting to notice that the CGHS model does {\it not}
belong to this set. A further remark is the fact that if one
requires the renormalizability of the theory in the usual sense,
then one can renormalize the metric in the theory (39) through the
following transformation
\beq
g_{\mu\nu} = \exp \left[ \frac{1}{\epsilon} \left( \frac{1}{C}+
\frac{Z}{2{C'}^2} \right) \right] g_{\mu\nu}^R.
\eeq
Then, the 2D dilaton-Maxwell theory (39) is multiplicatively
renormalizable off-shell in the usual sense for the following
choice of potentials:
\beq
V= \exp \left[ aC + \int \frac{Z\, d\Phi}{2C'}\right], \ \ \ \ \
f= \exp \left[-bC - \int \frac{Z\, d\Phi}{2C'} \right],
\eeq
where $a$ and $b$ are arbitrary constants. For the theory (3) the
above potentials are of Liouville type, what is in full agreement
with refs. [7,14].
\vspace{5mm}

\noindent{\large \bf Acknowledgments}

We are grateful to I. Antoniadis, F. Englert, Y. Kazama, A.A.
Slavnov and I.V. Tyutin for useful discussions at different stages
of this work.
S.D.O. wishes to thank the Japan Society for the Promotion of
Science (JSPS, Japan) for financial support  and the
Particle Physics Group at Hiroshima University for kind
hospitality.
E.E. has been supported by DGICYT (Spain), research project
PB90-0022, and by the Generalitat de Catalunya.
\bigskip

\appendix

\section{Appendix}

In this appendix we will show that the results of sect. 3 actually give
also the one-loop counterterms in 2D $R^2$-gravity (for a discussion of
different models of such theory see [16-18]).

Let us consider 2D $R^2$-gravity as defined by the action
\beq
S=-\int d^2x \, \sqrt g \, \left( \Lambda - \frac{a}{4} \,
R^2 \right),
\label{a1}
\eeq
where $a$ and $\Lambda$ are dimensional parameters. One can rewrite
(\ref{a1})
 by introducing the auxiliar scalar field (dilaton), as in [18],
\beq
S=-\int d^2x \, \sqrt g \, \left( R \Phi + \frac{1}{a} \, \Phi^2 +
\Lambda \right).
\label{a2}
\eeq
Theories with the actions (\ref{a1}) and (\ref{a2}) are classically
equivalent. The theory given by (\ref{a2}) belongs to class (1), with
\beq
Z=0, \ \ \ \ \ C= \Phi, \ \ \ \ \ V=\Lambda
+ \frac{1}{a} \, \Phi^2.
\eeq
The one-loop effective action for this theory has been actually
calculated in sect. 3. The result is
 \beq
\Gamma_{div} = -\frac{1}{2\epsilon} \int d^2x \, \sqrt g \, \left[ 4 R
+\frac{2}{\Phi} \left( \Lambda  + \frac{1}{a} \, \Phi^2 \right)
 + \frac{4}{a} \, \Phi
 + \frac{1}{\Phi} \, \Delta\Phi
 - \frac{3}{\Phi^2} \, (\nabla^\lambda \Phi ) (\nabla_\lambda \Phi )
\right].
\eeq
Hence, we see that, with our procedure, we are able to calculate the
one-loop effective action explicitly
in some version of 2D $R^2$-gravity. It is also interesting to notice
that $\Gamma_{div}$ is on-shell finite too.

\end{document}